Alireza Ranjbaran
Ranjbarana@cardiff.ac.uk
Cardiff University School of Biomedical and Life Sciences

Azadeh Nazemi
azadeh1972@gmail.com
Perth, Western Australia
0000-0002-1138-309X

`

# A survey on Organoid Image Analysis Platforms

## Abstract

An in-vitro cell culture system is used for biological discoveries and hypothesis-driven research on a particular cell type to understand mechanistic or test pharmaceutical drugs. Conventional in-vitro cultures have been applied to primary cells and immortalised cell lines plated on 2D surfaces. However, they are unreliable in complex physiological environments and can not always predict in-vivo behaviour correctly. Organoids are multicellular spheroids of a primary donor or stem cells that are replaced in vitro cell culture systems and are widely used in biological, biomedical and translational studies. Native heterogeneity, microanatomy, and functionality of an organ or diseased tissue can be represented by three-dimensional in-vitro tissue models such as organoids. Organoids are essential in in-vitro models for drug discovery and personalised drug screening. Many imaging artefacts such as organoid occlusion, overlap, out-of-focus spheroids and considerable heterogeneity in size cause difficulty in conventional image processing. Despite the power of organoid models for biology, their size and shape have mostly not been considered. Drug responses depend on dynamic changes in individual organoid morphology, number and size, which means differences in organoid shape and size, movement through focal planes, and live-cell staining with limited options cause challenges for drug response and growth analysis. In addition, organoids are imaged on a vast range of platforms in large structurally complex phenotypes. Hundreds of organoid cultures generate high-volume images at once, which are challenging to inspect and interpret. Therefore, an automated coding-free, intuitive and scalable image analysis solution is required. This study primarily introduces the importance of the role of the organoid culture system in different disciplines of medical science and various scopes of utilising organoids. Then studies the challenges of operating organoids, followed by reviewing image analysis systems or platforms applied to organoids to address organoid utilising challenges.

**Keywords:** High throughput, Fully automated, Image analysis system, Organoids, Pharmaceutical drug testing

## Introduction

This section briefly introduces utilising organoids in different areas such as organ studies, investigating physiology, drug testing, cancer biology, precision therapeutics development, antibody detection, quantifying cellular antigens, responses to chemotherapies and chemoradiation, human brain development and dysfunction. Organoids are multicellular spheroids of a primary donor or stem cells. Native heterogeneity, microanatomy, and functionality of an organ or diseased tissue can be represented by three-dimensional in-vitro tissue models such as organoids. Organoids are required in in-vitro disease models for drug discovery and personalised drug screening. According to (Clevers, H et al., 2016), (Schweiger P. J et al.,2016), (Bredenoord, A. L. et al.,2017), and ( Dutta, D. et al.,2017), In- vitro culture systems have been replaced with using organoid cultures systems. Many researchers studied organoids of different organs, as Table I indicates:

Table I. Researches and organoid of Organs

| Research | Organ |
|---|---|

| Cruz-Acuña, R. et al.,2017 Múnera, J. O, et al.,2017; | Gut |
| --- | --- |
| Broutier, L. et al.,2016; Grapin‑Botton, A., 2016 and Kim, Y.et al,2016 | Pancreas |
| Serruya, M. D,2017 | Brain |
| Skardal, A., et al., 2015 | Liver |
| Turco, M. Y.,et al,2017 | Endometrium |

According to (Shamir, E. R. et al.,2014) and (Skardal, A. et al.,2016), organoids are a suitable model system for understanding development, investigating physiology, and drug testing. Studying cancer biology and precision therapeutics development utilises patient-derived tumour organoids (TO) as high-fidelity models. (Larsen et al., 2021). For more than 20 years, oncology therapy has used precision medicine or individualised treatment approaches which means identification therapy for the patient's disease's unique biology, as Table II indicates:

Table II. Researches and organoid of cancer

| **Research** | **Cancer type** |
| --- | --- |
| Gao et al., 2014 | Prostate |
| Van de Wetering et al., 2015 | Colon |
| Boj et al., 2015 | Pancreatic |
| Huang et al., 2015 | Pancreatic |
| Fujii et al., 2016 | Colon |
| Broutier et al., 2017 | Hepatobiliary |
| Turco et al., 2017 | Endometrial |
| Sachs et al., 2018 | Breast |
| Tiriac et al., 2018; | Pancreatic |
| Kijima et al., 2018; Nanki et al., 2018 | Esophagogastric |
| Romero-Calvo et al., 2019 | Pancreatic |
| Sachs et al., 2019 | non-small-cell lung cancer (NSCLC) |
| Boretto et al., 2019 | Endometrial |

(Ferguson et al., 2020) hired TOs for drug development and precision medicine studies. They discovered responses to chemotherapies by observing patient-derived TOs' mimics. (Ganesh et al., 2019) investigated chemoradiation response.

Brain organoids from Human pluripotent stem cells are self-organised into cytoarchitectural structures. The three-dimensional models allow studying human brain development and dysfunction without considering characterisation, spatial information for single-cell, histological analysis, and whole-tissue analysis(Albanese A. et al., 2020).

Native heterogeneity, microanatomy, and functionality of an organ or diseased tissue can be represented by three-dimensional in-vitro tissue models such as organoids. Organoids are essential in in-vitro models for drug discovery and personalised drug screening. Many imaging artefacts such as organoid occlusion, overlap, out-of-focus spheroids and considerable heterogeneity in size cause difficulty in conventional image processing. Despite the power of organoid models for biology, their size and shape have mostly not been considered. Drug responses depend on dynamic changes in individual organoid morphology, number and size. It means differences in organoid shape and size, movement through focal planes, and live-cell staining with limited options cause challenges for drug response and growth analysis. In addition, organoids are imaged on a vast range of platforms in large structural complex phenotypic. Hundreds of organoids cultured generate high-volume images at once, which are challenging to inspect and interpret. About 75% of samples in clinical trials have failed due to safety and efficacy issues. Hence, a more accurate screening approach is required to improve sample translation. As a preclinical model, drug discovery pipelines are affected by patient-derived organoids (PODs). Organoid size in various intra and inter-patient, cellular heterogeneities, temporal response, and phenotypic evaluation has some drawbacks (Spiller E. R. et al., 2021). Due to a lack of sufficient knowledge of differential cellular states or genetic backgrounds, studies indicate different reactions to targeted therapies for tumours.

In 2017 Juan C Caicedo et al. recommended a method for converting the collection of microscopy images to high-quality image-based (e.g. morphological) profiles to pursue biological discovery, experimental designs, and laboratory preferences.

In 2017 Andrew M. K. Law et al. designed Andy's Algorithms to perform batch-processing of 3,3′-diaminobenzidine (DAB) immunohistochemistry, proximity ligation assays (PLAs) and other standard assays in a quick, accurate and repeatable manner.

In 2018, Borten et al. developed OrganoSeg to segment, filter and analysis 3D brightfield images. OrganoSeg quantifies brightfield phenotypes, providing insight into organoid regulation's /molecular and multicellular mechanisms of organoid regulation.

In 2019, Timothy Kassis et al. implemented OrgaQuant, to locate and quantify the size distribution of human intestinal organoids in bright field images using a deep convolutional neural network. The OrgaQuant model does not require any parameters and can analyse thousands of images without human interaction.

In 2020, Alexandre Albanese et al. presented the SCOUT pipeline for intact cerebral organoids with an automated multiscale comparative analysis to clear, label, and image intact organoids.

In 2021, Erin R. Spiller et al. proposed an image-based approach and high-content assay to obtain object-level information on 3D patient-derived tumour organoids without vital dyes requirements. This method tracks the dynamic response of individual organoids to various drugs in a robust, nondestructive manner utilising brightfield images at different time points. In addition, they developed a web-based open-source data visualisation system to simplify the analysis of extensive complex data. This research proved that imaging, computer vision, and machine learning are practical in biological research.

In 2021, Brian M. Larsen et al. utilised chemically defined media optimised from over 1,000 patients to describe the robust pan-cancer TO platform to predict heterogeneity in drug responses for solid cancers. They used tumour genetic and transcriptomic concordance to accelerate the broad implementation of organoid models using molecular data, precision medicine research, personalised therapeutic profiling programs, and defined minimal media for organoid initiation and propagation.

Gritti N. et al. 2021 developed MOrgAna, a python-based platform using machine learning to segment, quantify and visualise morphological features of high-volume organoid images at once.

In 2022, Jonathan Matthews et al. developed OrganoID, a robust image pixel-by-pixel analysis platform, in brightfield and phase-contrast microscopy experiments to automatically recognise, label, and track single organoids.

The next section reviews different image analysis platforms or software packages for organoid image processing and compares the methods and performance.

**Search StrategyMethods:**

Several scoping searches were conducted before performing a literature search to provide an overview of the literature. It was realised during these searches that there were many image analysis systems used for organoid studies. To ensure no relevant papers were excluded, a search was conducted to collect as many systems as possible before deciding on the search terms. This allowed the final search to be broad enough to include all systems from 2003 until now.

After conducting scoping searches, the research question was proposed to collect a survey of image organoid image analysis systems addressing drawbacks in using organoids. This was done to ensure that the research has survey criteria and meets the FINER (Feasible, Interesting, Novel, Ethical, Relevant) requirement. All relevant repositories were searched for corresponding literature dating back to 2003. As the research objective is to find a fully automated, high throughput and general-purpose image analysis system. Table III. indicates reviewed journals and their relevant database

Table III. reviewed journals and their relevant database

| Journal names | database |
|---|---|
| Elsevier | PubMed, MEDLINE, EMBASE, and Scopus |
| CellPress | PubMed, MEDLINE, and the Cochrane Library |
| BioRvix | PubMed, MEDLINE, EMBASE, and Scopus |
| Scientific Report | PubMed, Web of Science, and Scopus |
| Frontiers in Oncology | PubMed and Web of Science |
| Springer | PubMed, Web of Science, and Scopus |

The following section briefly explain about the medical databases used by this study.
**Pubmed**:
PubMed is a free search engine accessing the MEDLINE database of references and abstracts on life sciences and biomedical topics primarily. The United States National Library of Medicine (NLM) at the National Institutes of Health maintains the database as part of the Entrez system of information retrieval. PubMed indexes more than 28 million citations for biomedical literature from MEDLINE, life science journals, and online books. Citations may include links to full-text content from PubMed Central and publisher websites.

**EMBASE:**
EMBASE is a biomedical and pharmacological database that indexes journal articles and conference proceedings. It is produced by Elsevier and is available via the Elsevier products ScienceDirect and Scopus. EMBASE covers a wide range of topics, including drug discovery, pharmacology, toxicology, epidemiology, and healthcare. It contains over 22 million records, dating back to 1947. EMBASE is updated weekly and covers over 8,000 journals, making it an essential resource for researchers in the biomedical and pharmacological sciences.

**Web of Science:**
Developed by the Institute for Scientific Information, the Web of Science is a platform that allows users to access research data and citation information. The platform includes the Science Citation Index, the Social Sciences Citation Index, and the Arts & Humanities Citation Index. The Web of Science also provides access to the Conference Proceedings Citation Index and the Book Citation Index. The platform is designed to help users identify, evaluate, and use scholarly research data.

**Scopus:**
Scopus is a large abstract and citation database of peer-reviewed literature: scientific journals, books and conference proceedings. Delivering a comprehensive overview of the world's research output in the fields of science, technology, medicine, social sciences, and arts and humanities, Scopus features smart tools to track, analyze and visualize research.

**Cochrane Library**
The Cochrane Library is a collection of six databases that contain different types of medical information. The databases are The Cochrane Database of Systematic Reviews (CDSR): This database contains reviews of healthcare interventions, such as drugs, medical devices, and surgeries. The Cochrane Central Register of Controlled Trials (CENTRAL): This database contains information on clinical trials. The Cochrane Methodology Register (CMR): This database contains information on methods used in healthcare research. The Health Technology Assessment Database (HTA): This database contains information on the effectiveness and safety of health technologies, such as drugs, medical devices, and diagnostic tests. The NHS Economic Evaluation Database (NHS EED): This database contains information on the cost-effectiveness of healthcare interventions. The Joanna Briggs Institute EBP Database (JBI): This database contains information on evidence-based practice.

**Result**
This section presents a summary review of some image analysis pipelines for organoids. Table IV denotes these platforms, scopes, techniques and properties in terms of highlighted keywords for this research

Table IV. Image analysis platform for organoids

| Platform | Applied to/purposes | Technique | Batch processing/High throughput | Fully automated |
|---|---|---|---|---|
| Metamorph | Time-lapse, multi-dimensional acquisition and 3D reconstruction on microscopic image analysis of live cells | Image processing | Not tell<br>Manual user interface | Yes |

| Caicedo | Biological systems on a large scale by using chemical and genetic perturbations. | Data-analysis strategies for image-based cell profiling | Yes | Yes |
|---|---|---|---|---|
| CALYPSO | Intrinsic heterogeneity of cancer tissues | Image processing, | Yes | Not tell |
| Andy's Algorithm | Oncology drug development | Image processing | Yes | Yes |
| OrganoSeg | Brightfield 3D organoid populations | Morphometry | Standalone Not say | Yes |
| OrgaQuant | Human Intestinal Organoid Localization and Quantification | Faster R-CNN | Not tell | Yes |
| SCOUT | 3D phenotyping of human cerebral organoids | UNet | Yes | Yes |
| Larsen | Remove vital dye requirements. | Pix2Pix GAN | Yes | Yes |
| OBIA | Drug treatment responses on Patient-Derived Organoids (PDOs) | Supervised linear classifier to separate live and dead | Yes | Yes |
| MOrgAna | Segmentation and quantifying organoid brightfield images. | Gaussian /Laplacian filters, Logistic regression | Yes | Yes |
| OrganoID | Pancreatic ductal adenocarcinoma (PDAC) | modified UNet | Yes | Yes |

**Literature review**

Utilising organoids assists scientists in different areas such as organ studies, investigating physiology, cancer biology, precision therapeutics development, responses to chemotherapies and chemoradiation. Organoids can be imaged on platforms such as benchtop stereoscopes or high-content confocal-based. Investigation, interpretation and inspection of high-volume organoids are problematic. Data collection from various time points and conditions among thousands of regions of interest is still challenging for biologists. Using organoids still faces some drawbacks that need to be investigated. Some problems regard ignorance of organoid size, shape, and morphological status. Furthermore, images have been provided in a high volume in quantity requiring automatic batch processing. Therefore, Fully automated high throughput image analysis pipelines for organoids are required. Image analysis is a powerful tool in many medical disciplines. Today, researchers use machine learning to improve image processing results. The software using machine learning are MetaMorph, Imaris, Harmony, ZEN, FIJI, CellProfiler and ilastik (Berg et al., 2019; Schindelin et al., 2012; Carpenter et al., 2006). Some identify thousands of cells in vast fields and extract biologically related features (McQuin et al., 2018).

**Metamorph(2003)**

MetaMorph is a microscopic image analysis tool to manipulate various image processing tasks such as intensity logging, advanced morphology, colocalisation, image stack alignment, montage generation, movie making, 3D rendering, image colour combination, light equalisation, topographical surface map generation, convolve and deconvolve images, arithmetic operations, orthogonal planes visualisation, image stitching and 3D calculation.

The graphical interface offers colour combination, integrated morphometry analysis, region measurements and counting cells. Metamorph has many optional modules such as counting nuclei, neurite outgrowth analysis, tracking motile cells or assessing cell cycle phases.

MetaMorph software is a quick, practical and robust system. It performs time-lapse and multi-dimensional acquisition and 3D reconstruction.

Furthermore, Methamorph has brightening operations for biological experiments applying to live cell imaging.

**Caicedo research**

In 2017, Juan C Caicedo et al. recommended a method for converting the collection of microscopy images to high-quality image-based (e.g. morphological) profiles to biological discovery, experimental designs, and laboratory preferences.

Image-based cell profiling is a high-throughput microscopy system for quantifying phenotypic differences among various cell populations. It uses chemical and genetic perturbations and provides an approach to studying biological systems on a large scale (Caicedo, J.C et al., 2017).

Their image analysis system performs two main steps:

1-Illumination detection and correction using retrospective multi-image. In the centre of the field of view, pixels are brighter than those on the edges.

2-Pixel foreground or background classification using image segmentation.

Figure. 1 illustrates the main Caicedo steps

> Image illumination detection and correction --> Pixel foreground or background classification --> Image segmentation

Figure.1 Caicedo overview

**Comprehensive image analysis procedure for structurally complex organotypic cultures (CALYPSO)**

In 2017, Anne-Laure Bulin et al. proposed a comprehensive high-volume image analysis system. They aimed to study complex heterotypic organoids to discover the therapeutic efficacy architecturally. They extracted fluorescence intensity-based parameters for tumour treatment. CALYPSO measures the organoid area without considering the shape and volumetric estimations. For binarization, adaptive thresholding was used instead of a constant threshold for accurate tumour nodules detection. To eliminate out-of-focus objects, filtering has been done by adding fluorescence channels and binarising Otsu's threshold method. The obtained binary images were multiplied with initial masks.

Organoids are indexed based on final masks. These indices are unique for each organoid. CALYPSO calculates tumour nodules viability factor using fluorescence intensity. This factor is varied between zero and one and indicates organoids' health.

CALYPSO performed the following steps for image analysis: masking, background subtraction, threshold calculation and finally, live area thresholding. Viability and live area fraction are used for correlated analysis. Live area fraction is obtained from the rate of the fraction of live area to total area. Viability is obtained from the rate of live intensity to the total value of live intensity and dead intensity.

Figure.2 illustrates the main steps of CALYPSO

> Adaptive threshold --> Filtering --> Adding fluorescence channels --> Otsu threshold

Figure.2 CALYPSO overview

**Andy's Algorithm**

Image analysis based on antibody detection faces many challenges in quantifying cellular antigens and obtaining accurate high-throughput quantitation, specifically for users with insufficient image processing skills. Cognitive bias and scientific reproducibility are emerging problems in oncology drug development.

In 2017 Andrew M. K. Law et al. designed Andy's Algorithms to address the stated issues. This algorithm performs batch-processing of 3,3′-diaminobenzidine (DAB) immunohistochemistry, proximity ligation assays (PLAs) and other standard assays in a quick, accurate and repeatable manner.

Andy's algorithm performs optimisation before image processing to determine optimal parameters. These parameters are used to address the problems caused by variation of cell types, tissue type, fixation method, antigen detection method, staining pattern, colour discrimination, magnification, illumination settings, and resolution.

Andy's algorithm provides interactive environments for users with basic knowledge. It was designed for compassionate pre-clinical oncology purposes. Andy's algorithm is a high throughput and iterative image analysis platform for biology, this platform does not require advanced image processing skills.

**OrganoSeg**

In 2018, Borten et al. developed OrganoSeg as an open source and parameters based on conventional image processing techniques to segment, filter and analyse 3D brightfield images. OrganoSeg quantifies brightfield phenotypes, providing insight into the molecular and multicellular mechanisms of organoid regulation. Its high performance has been approved by classifying 5167 breast-cancer spheroids and 5743 colon cancer cells.

OrganoSeg works on jpg, png or tiff format, grayscale images, performs an open-close morphological function, smooths bulk components preserving sharp spheroid boundaries, and then binarises the smoothed image by local adaptive thresholding or Otsu's method. Users are enabled to find minimum threshold size and separate poor growth organoids by segmentation. OrganoSeg extracts up to 23 morphometry standard measures for downstream. It works with 3D-cultured breast images or movies, brightfield, phase-contrast, and differential interference contrast images. OrganoSeg, regardless of confocal or high contrast fluorescence micrographs, translates obtained images to measurable datasets. High volume organoid shape analysis generates a collection of non-continuous morphological references. Combining OrganoSeg and RNA sequencing indicated a relationship in colorectal cancer. OrganoSeg is standalone with a graphical user interface to quantify organoid cultures and 3D spheroid transmitted-light images.

Figure.3 illustrates the main OrganoSeg steps.

Grayscale-->Open/close morphology-->Adaptive Thresholds-->Superimpose-->Binarise-->Filter noise-->Fill holes-->Smooth-->Identify ROI-->Extract metrics

Figure.3 OrganoSeg overview

**OrgaQuant**

In 2019, Timothy Kassis et al. described OrgaQuant implementation, a deep convolutional neural network to locate and quantify bright field human intestinal organoids images distribution size. To train OrgaQuant, they manually annotated human intestinal organoid images and generated a dataset. They conducted this research to address the lack of technique to localise and quantify organoids in a 3D environment and the presence of many imaging artefacts (Kassis T. et al., 2019). They used extracted features from a neural network (NN) for Clustering or subtle changes detection in organoid morphology.

Object detection is a complicated process in computer vision, particularly in cases with speed concerns. Some developed detection algorithms have been designed to be quick and accurate such as single Shot Multibox Detector (SSD) and You Only Look Once (YOLO). OrgaQuant was designed based on the Region Convolutional Neural Network (R-CNN) and or faster R-CNN. Faster R-CNN uses ResNet and Inception architecture.

OrgaQuant for pre-training utilises COCO box annotations and needs an extensive dataset provided by augmentation. OrgaQuant automatically localises an organoid within a brightfield image and annotates the bounding box. Downstream image processing uses cropped organoid images. This pipeline measures the intestinal organoid size.

OrgaQuant facilitates tracking organoid growth kinetics in droplets during the time. OrgaQuant is a basic, intelligent and open-source organoid quantification technique. an open source and parameters based on conventional image processing techniques.

The OrgaQuant model does not require any parameters and can analyse thousands of images without human interaction. Figure.4 illustrates the main OrgaQuant steps

**Brightfield image --> Faster R-CNN --> Localised organoid**

Figure.4 illustrates the main OrgaQuant steps

**Single-cell and Cytoarchitecture analysis of Organoids using Unbiased Techniques (SCOUT)**

In 2020, Alexandre Albanese et al. presented the SCOUT pipeline for intact cerebral organoids with an automated multiscale comparative analysis to clear, label and image organoids. From the microscopic fluorescence dataset, SCOUT CNN extracts hundreds of features characterising molecular, cellular, spatial, cytoarchitectural, and organoid-wide properties. SCOUT performed a comprehensive analysis of 46 intact organoids and ~100 million cells and revealed quantitative multiscale "phenotypes" for organoid development and Zika virus infection.

Imaging each organoid takes about 15 minutes for reliable, accurate single-cell analysis and 300 multiscale features takes about six hours.

SCOUT has performed quantifying spatial features such as cellular context, ventricle morphology, cytoarchitecture distribution, and uncommon events detection. Zika virus infection decreases cell population, and this pipeline quantifies this reduction.

SCOUT has three main modules: single cell analysis, regional architecture analysis, whole organoid analysis, and finding correlation.

UNet was slightly modified, and two layers were added before and after the UNet bottleneck to improve the model. Then modified UNet was trained using weighted binary class entropy (WBCE). WBCE support model to converge to accurate ventricle segment. This model was used to segment ventricles automatically. Experimental results denote 97.2% Dice score similarity for UNet. Quantifying multiscale feature correlation, maturation-related changes, protocol comparisons, and

Zika virus pathology demonstrated SCOUT's power. SCOUT is a versatile platform for automatic single-cell analysis.

Figure.5 illustrates the main steps of SCOUT

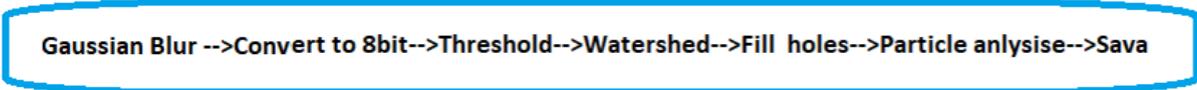

Figure5. SCOUT overview

**Larsen**

In 2021, Larsen et al. proposed a pan-cancer precision medicine organoid platform. Genomic/transcriptomic fidelity has been extracted from over 1000 patients' tumour organoid(TOs) cultures. They determined solid tumour chemical minimal media. They studied growth factors to initiate and propagate organoids. Their results indicated that Epidermal Growth Factor (EGF) and Noggin are sufficient for most organoid cultures. They developed a convolutional neural network(CNN) based on pix2pix Generative Adversarial Network (GAN) and predicted label-free light microscopy drug response. CNN contains a fluorescence generator and a viability discriminator. Pix2pix GAN accuracy has been approved in cell biology by Isola et al. 2017; Tsuda and Hotta, 2019. They adopted 70x70 patch GAN as a viability discriminator (Isola et al., 2017). Concatenated brightfield and fluorescent are inputs of the discriminator. The generator gets a brightfield image as input and generates fluorescence.

Realistic image prediction and image translation from the black and white domain to the colourful domain is possible by Pix2Pix (Isola et al., 2017), conditional generative adversarial networks (GANs), and CycleGAN (Zhu et al., 2017).

This technique was developed as a light-microscopy-based assay to omit costly vital dye requirements. High content image analysis was performed by inverting confocal microscopy images, followed by ability measurements. Then binary classification for TO viability was executed for drug screening optimisation. Figure.6 illustrates the main Larsen steps

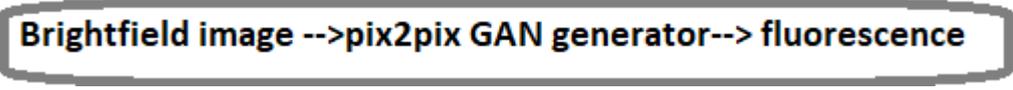

Figure.6 Larsen steps overview

**Object-Based Image Analysis (OBIA)**

In 2021, Erin R. Spiller et al. developed an object-based image analysis (OBIA) system to convert cells to population-level analysis and investigate disturbances in drug treatment responses of heterogeneous object-based Patient-Derived Organoids (PDOs). It observes phenotypic changes. OBIA considered the textural and morphological features from brightfield images as discriminators between dead and live organoids. They used a supervised linear classifier to separate live and dead organoids for drug response recognition. The OBIA ML image analysis was initially performed in a small sample set. They designed a tool to visualise the differences of all collected features over time. The OBIA team developed a web-based Application Programming Interface (API) to upload machine learning output analysis and download feature metrics and survival curves.

They proposed a label-free high content screening (HCS) robust, on-time approach working with colorectal cancer (CRC) organoid live-cell image. It is an automated dynamic cellular visualisation and multiparametric extraction data system.

Moreover, this method can be scaled to apply to high-volume drug screen data of different cancer types PDOs. OBIA has done computational intensive image analysis automatically and iteratively. Figure.7 illustrates the main OBIA steps

> Brightfield images --> Textural and morphological features --> Supervised linear classifier --> Classified live and dead organoids

Figure.7 OBIA overview

**MOrgAna**
Gritti N. et al. 2021 developed MOrgAna, a python-based platform using machine learning to segment, quantify and visualise morphological features of high-volume organoid images at once. MOrgAna is an appropriate package for users with basic to advanced skills.
So, this technology assists in atlas-free 3D whole-organoid analysis to gain quantitative comparative analysis between experimental conditions. Brightfield image availability, in contrast, fluorescence is the reason behind utilising them. MOrgAna monitors the organoid in bright field mode for biomedical purposes. The training dataset includes morphological parameters, Gaussian and Laplacian filters, Logistic regression, improved flexibility and feasibility. This pipeline is designed on a multilayer perceptron for organoid recognition.
Figure.8 illustrates the main MOrgAna steps

> Brightfield image --> morphological parameters --> Gaussian and Laplacian filters --> Logistic regression

Figure.8 MOrgAna overview

**OrganoID**
In 2022, Jonathan Matthews et al. developed OrganoID, a robust image pixel-by-pixel analysis platform, in brightfield and phase-contrast microscopy experiments to automatically recognise, label, and track single organoids. This model was trained based on pancreatic cancer organoid images and tested on images of pancreatic, lung, colon, and adenoid cystic carcinoma organoids. OrganoID is an open-source image analysis platform which determines organoid morphology pixel-wised. This platform does not require fluorescence or transgenic labelling. It can analyse different organoid types in microscopy experiments promptly. It is designed based on a modified UNet and has minimum layers for quick performance. The training dataset was pancreas cancer samples. Testing validating images come from pancreatic, lung, colon, and adenoid cystic carcinoma with more than 89% tracking accuracy for four days of the validation period. OrganoID performs two highlighted tasks: individual organoid tracking over time and Drug screening experiments. General image processing steps with OrganoID are contour detection, single and bulk organoid separation and determination of response over time by time-lapse image sequence. CNN has been designed to indicate the presence of organoid probability in each pixel. UNet architecture includes layers of convolutional, maximum filtering, concatenation, extraction and localisation features.
The training dataset was obtained from 66 manually organoid microscopy images of pancreatic ductal adenocarcinoma (PDAC), then collected samples divided into 80% training and 20% validating and augmentation applied to the training dataset. The OrganoID platform effectively and accurately uses organoid images in high-volume data physiological research.

Figure.9 is an illustration of the main steps of OrganoID

> Neural Network predication map-->Canny edge detection--> Watershed

Figure.9 OrganoID overview

**Critical Analysis**

Reviewing image analysis pipelines for organoids of 11 researches indicates that since 2019 four studies have used deep learning for organoid analysis. Generally using deep learning, in contrast, traditional methods require larger training datasets, which is expensive, time-consuming, and requires expert annotation. Although such training dataset preparation needs more effort, the result of models will be more robust and efficient.

Almost all pipelines are automated, and more than 75% of them are high throughput and can process a high number of images at once.

As Table IV indicates, these pipelines have been designed for different targets. Since five pipelines specifically have been designed for oncology studies; thus it denotes organoid significance in cancer studies. Almost all projects consider organoids' morphology for different purposes, such as drug response monitoring. They address the issues regarding ignorance of organoid shape and size.

**Conclusion and future development**

In this study, we have introduced the vast range of organoids usage in biology and medical research, such as organ studies, investigating physiology, drug testing, cancer biology, precision therapeutics development, antibody detection, quantifying cellular antigens, responses to chemotherapies/chemoradiation, and human brain development /dysfunction.

Then briefly describe the drawbacks of using organoids and finally review eleven designed image analysis systems to solve the drawbacks. Since 2014, more than 23 researches have been conducted to observe different organs' functions and dysfunction using organoids.

Organoids can be imaged on platforms such as benchtop stereoscopes or high-content confocal-based microscopes. Investigation, interpretation and inspection of high-volume organoids are problematic. Data collection from various time points and conditions among thousands of regions of interest is still challenging for biologists. Some problems regard ignorance of organoid size, shape, and morphological status. Hence, the essential properties of an organoid image analysis pipeline are fully automated, high throughput, and considering morphology features. We reviewed the development of image analysis systems for Organoids in biological disciplines. Starting from 2003 with MetaMorph development which was able to perform some simple image processing tasks such as montage, rendering, alignment, colour combination and arithmetic operation robustly and quickly. Proceeding to 2017, Caicedo et al. developed another system to encourage biological observation and experimental laboratory preference, followed by the development of CALYPSO to measure the viability of an organoid to discover its therapeutic efficacy. At the same time, Andy's algorithm came as a high throughput, iterative and user-friendly image analysis biology platform for compassionate pre-clinical oncology purposes. In 2018, OrganoSeg was released as a standalone graphical high-volume platform to extract metrics from organoids. OrganoQuant in 2019 was designed based on faster RCNN to address the issues for organoid quantification regarding image artefacts. In 2020, SCOUT was revealed as a versatile platform based on UNet and came in 3 modules: single cell, regional and whole organoid to automatically segment ventricles. In 2021, Larcen's pan-cancer organoid platform was designed for precision medicine based on pix2pix GAN network image translation. This platform was developed to remove costly vital dye requirements and optimise drug screening. OBIA was released for drug response recognition levels by classifying dead and live organoids. The OBIA team also developed a visualisation tool and web-based API to address the biologist's challenges regarding high-volume data. MOrgAna was developed for organoid segmentation, quantification and visualisation. OrganoID was designed for organoid tracking and drug screening based on UNet.

Eleven projects have been studied. These projects aimed to provide organoid image analysis systems.

Regardless of the area these pipelines are designed for, they all meet the general criteria for organoid image analysis systems. These criteria are being automated, high throughput and considering organoid morphology.

Four out of eleven pipelines used Deep Neural Network(DNN). DNN, despite being more expensive, expert-required annotation, time-consuming preparation generates reliable results. For instance, Figure 10 illustrates the results of computed tomography (CT) of abdominal multi organs image segmentation using UNet (we developed the UNet model prior to this research to evaluate the segmentation accuracy of this model). This figure shows spleen, right kidney and left kidney segmentation from top to bottom. As observed, segmentation accuracy decreased from top to bottom due to reducing the number of images in the training dataset.

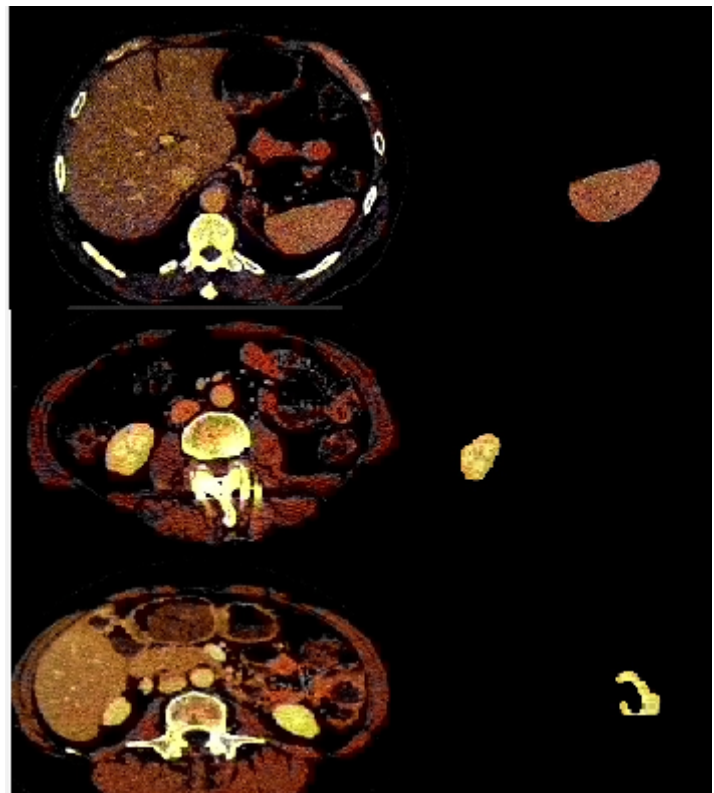

Figure. 10 UNet abdominal CT scan organs segmentation

As these pipelines followed different targets, comparing them in terms of performance can not be correct. We can only compare them based on image processing techniques or can compare pipelines with common targets/scops, as Table V indicates.

The summary of nine reviewed pipelines shows in Table V.

Table V. pipelines overview

| Pipelines | Initial Image processing steps |
|---|---|

| Caicedo | Image illumination detection and correction →pixel background/ foreground classification |
| --- | --- |
| CALYPSO | Adaptive threshold→ Filtering →Adding fluorescence channel→Otsu threshold |
| OrganoSeg | Grayscale→open/close Morphology→Adaptive threshold→Superimpose→Filter noise and holes→Smooth→ROI |
| OrgaQuant | Brightfield images→Faster RCNN→Localised organoid |
| SCOUT | Gaussian →8 bits→Threshold→Watershed→Fill holes→Particle analysis |
| Larsen | Brightfield images→Pix2pix GAN→generator→Fluorescence |
| OBIA | Brightfield images→ Morphological features→Linear classifier |
| MOrgAna | Brightfield images→ Morphological features→Laplacian and Gaussian filters→Logistic Regression |
| OrganoID | NN prediction map→Canny edge→ Watershed |

**Future work and further development**

Since during this study, despite a limited training dataset, we successfully developed UNet to be familiar with the NN model design. We could get reasonable cross-validation results; thus, further development will be conducted to design a trustworthy system to classify live and dead cells using Pix2Pix GAN in the oncology area. This system includes two networks, a UNet as a generator and ResNet as a discriminator to classify gold standard and live/dead masks generated by the generator. The generator accepts brightfield images and generates live/dead masks. Two models will be trained by this method: one model generates masks of live cells, and another one generates masks of dead cells. For providing training datasets for two models, we can use CALYPSO or OBIA. These pipelines support us in saving time and budget. Then brightfield images and dead/live organoid masks will be concatenated as input and output of the generator. To have samples for testing or cross-validation training, datasets will be divided into 80% train and 20% test.

Reviewing table V indicates the ideal pipeline for organoid image analysis is a fully automated, high-throughput and general-purpose system designed based on Deep Neural Network, covering brightfield and fluorescence images.